\documentstyle[aps,twocolumn]{revtex}

\input psfig

\begin{document}

\title{Comment on ``Quantum Melting of the Quasi-Two-Dimensional Vortex 
Lattice in $\kappa-$(ET)$_2$Cu(NCS)$_2$''}  
\author{Y. Kopelevich$^+$ and P. Esquinazi$^*$}\address{$^+$Instituto
de Fisica,  Unicamp, 13083-970 Campinas, Sao Paulo, Brasil}  
\address{$^*$Abteilung Supraleitung und  Magnetismus, Universit\"at
Leipzig, Linn{\'e}str. 5, D-04103 Leipzig, Germany} 

\maketitle 
\pacs{74.70.Kn,74.25.Ha,74.60.Ec,74.60.Ge}
\vspace{-13mm}
In a recent Letter Mola et al. \cite{mola} reported magnetization 
measurements $M(H,\theta)$ performed on the organic superconductor $\kappa-$(ET)$_2$Cu(NCS)$_2  (T_c = 
9.1~$K) as a function of the magnetic field $H$ applied at different angles $\theta$ with respect to the 
$a$-axis direction. The results \cite{mola} demonstrate:
 (a) the occurrence 
of pronounced irreversible magnetization $M_{\rm irr}(H)$ jumps and (b) their sudden 
cessation for $H \ge H_m(T,\theta)$. 
The boundary line $H_m(T)$  has 
been interpreted by Mola et al. 
as the Q2D vortex lattice (VL) quantum melting phase transition line \cite{mola}.
The purpose of this comment is to show that the results 
can be understood in a simple way without invoking "quantum melting 
phase transition".

As has been recently  reported in detail \cite{esq}, the termination of  flux jumps with field accompanied 
by a kink -or second magnetization peak (SMP)-like- anomaly in $M_{\rm irr}(H)$  is also observed in 
conventional Nb superconducting films and can  be naturally understood within the framework of the 
thermomagnetic flux-jump instability theory (TMI) \cite{min} without invoking  VL 
transformations. Using both global and local magnetization $M(H)$ 
measurements, it has been demonstrated that at low enough temperatures 
and at $H < H_{\rm SMP}(T)$ TMI leads to the reduction of  $|M_{\rm irr}(H)|$ 
relative to the isothermal critical state 
value, and this is reestablished at $H \ge H_{\rm SMP}(T)$ due to the decrease of 
the critical current density $j_c$ 
with field \cite{esq}. It has been shown also \cite{esq} that the $H_{\rm SMP}(T)$ 
can be approximated by the 
equation $H_{\rm SMP}(T) = H_0(1-(T/T_0)^2)$, where $H_0$ is a free parameter 
and $T_0 < T_c$ depends on the 
intrinsic pinning properties as well as on the effective 
film size $s_{\rm eff} \sim (wd/2)^{1/2}$ ($w$ and $d$ being the film width and 
thickness) of the sample. 

Figure 1 shows the ``melting" line $H_m(T/T_0)$ obtained in \cite{mola} 
for a  crystal  with dimensions $l \times w \times d = 1 \times 1 \times 0.3~$mm$^3$. 
To facilitate the comparison with the
 Nb films we have normalized the temperature scale by the value $T_0 = 0.152~$K. 
As can be seen in Fig. 1, the $H_m(T/T_0)$ line  is similar to the
$H_{\rm SMP}(T)$ lines obtained in Nb films. As for the Nb films, $H_m(T)$ can be very well fitted 
by the equation $H_m(T) = H_0 (1-(T/T_0)^2)$
 with $\mu_0 H_0 = 3.8~$T and $T_0 = 0.152~$K, whereas $T_0$ for the Nb films runs from $\sim 4~$K to 
$\sim 7~$K depending on the effective size of the films \cite{esq}. 
The difference in $T_0$ between
the two superconductors is mainly due to the difference in the magnitude of pinning. 
The TMI theory \cite{min} predicts a critical 
effective sample size $s_{\rm crit} = 10H_{fj}/4\pi j_c$ below which no flux jumps appear. Here $H_{fj}$ is the 
field at which first flux jump occurs. Taking $H_{fj} \sim 1-2~$kOe and in-plane $j_c(0) 
\sim 4 \times 10^4~$A/cm$^2$ \cite{mola,mola2}
one gets $s_{\rm crit} \sim 0.2 - 0.4~$mm which coincides with the effective crystal size $s_{\rm eff} \sim 
0.4~$mm, implying that the instability criterion ($s_{\rm crit} < s_{\rm eff})$ can be reached at very 
low temperatures $(T_0/T_c << 1)$ as the experiment in  \cite{mola} demonstrates. 

\begin{figure}
\centerline{\psfig{file=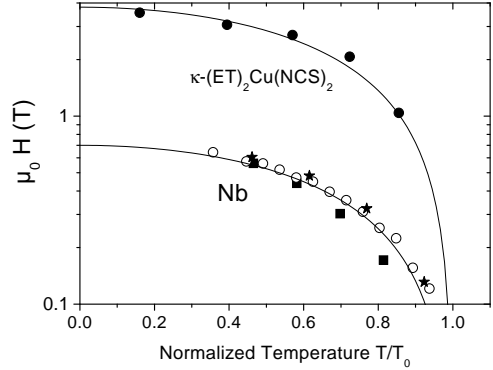,height=2.2in}}
\caption{$(\bullet): H_m(T)$ data points from \protect\cite{mola}. The solid lines correspond to the 
equation $H(T) = H_0(1- (T/T_0)^2)$. $\mu_0 H_0 = 3.8~$T and $T_0 = 0.152~$K ($\mu_0 H_0 = 
0.7~$T and $T_0 = 4 \ldots 7~$K) 
for the upper (lower) data points. The lower data
points correspond to $H_{\rm SMP}(T)$ obtained for Nb films \protect\cite{esq}.} 
\end{figure}
The increase of $H_m$ with $\theta$ reported in \cite{mola} can be understood as follows. First of all, 
note that within the TMI approach the expected decrease of  $H_m$ with $\theta$ due to reduction 
of  $s_{\rm eff}$  from $\sim 0.4~$mm to $s_{\rm eff}\sim d = 0.3~$mm \cite{swa}
 should be negligible. However, the activation energy of pancake vortices increases with $\theta$ \cite{kop}. 
Therefore, a larger applied field is needed to stabilize the isothermal critical state \cite{mch}. We note 
further that the coincidence of a single $H_m(\theta)$-data point obtained for the second (larger) 
crystal with the reported $H_m(\theta)$ \cite{mola} can be merely due to a difference in $j_c$.
In summary the $H_m(T, \theta)$ obtained in \cite{mola} is simply the
boundary in the $H-T$ plane that separates the TMI regime from the isothermal 
Bean-like critical state.


\begin{references} 
\bibitem{mola} M. M. Mola et al., Phys. Rev. Lett. {\bf 86}, 2130 (2001).
\bibitem{esq} P. Esquinazi et al., Phys. Rev. B {\bf 60}, 12454 (1999) and references
therein.
\bibitem{min}R. G. Mints and A. L. Rakhmanov, Rev. Mod. Phys. {\bf 53}, 551 (1981) and references 
therein.
\bibitem{mola2}M. M. Mola et al., cond-mat/0011227.

\bibitem{swa}P. S. Swartz and C. P. Bean, J. Appl. Phys. {\bf 39}, 4991 (1968).

\bibitem{kop}Y. Kopelevich et al., Physica C {\bf 183}, 345 (1991).

\bibitem{mch}M. E. McHenry et al., Physica C {\bf 190}, 403 (1992).

\end{references}
\end{document}